\documentclass[aps,prx,twocolumn,showpacs,superscriptaddress,longbibliography]{revtex4-2}
\usepackage{amsmath,amssymb,graphicx}
\usepackage{graphicx,epstopdf}
\usepackage{gensymb}
\usepackage[dvipsnames]{xcolor}
\usepackage[T1]{fontenc}
\usepackage[utf8]{inputenc}
\newcommand{\be}{\begin{equation}}
	\newcommand{\ee}{\end{equation}}
\newcommand{\bea}{\begin{eqnarray}}
	\newcommand{\eea}{\end{eqnarray}}
\newcommand{\bse}{\begin{subequations}}
	\newcommand{\ese}{\end{subequations}}

\usepackage{multibib}
\usepackage{color}
\usepackage[colorlinks,bookmarks=false,citecolor=darkblue,linkcolor=red,urlcolor=blue]{hyperref}

\definecolor{darkred}{rgb}{0.7,0.0,0.0}

\definecolor{darkblue}{rgb}{0,0.02,0.45}

\definecolor{darkgreen}{rgb}{0.02,0.45,0.0}

\definecolor{violet}{rgb}{0.8,0.2,0.6}

\begin{document}
	
	\title{Static and dynamic properties of %Non-trivial magnetic order in 
		the frustrated spin-$\frac{1}
		{2}$ depleted-kagome antiferromagnet Cu$_7$(TeO$_3$)$_2$(SO$_4$)$_2$(OH)$_6$}
	\author{K. U. Akshay}
	\affiliation{School of Physics, Indian Institute of Science Education and Research, Thiruvananthapuram 695551, India}
	\author{Sebin J. Sebastian}
	\affiliation{School of Physics, Indian Institute of Science Education and Research, Thiruvananthapuram 695551, India}
	\affiliation{Ames National Laboratory, U.S. DOE, Iowa State University, Ames, IA 50011, USA}
	% \affiliation{Department of Physics and Astronomy, Iowa State University, Ames, IA 50011, USA}
	\author{Q.-P. Ding }
	\affiliation{Ames National Laboratory, U.S. DOE, Iowa State University, Ames, IA 50011, USA}
	% \affiliation{Department of Physics and Astronomy, Iowa State University, Ames, IA 50011, USA}
	\author{Y. Furukawa}
	\affiliation{Ames National Laboratory, U.S. DOE, Iowa State University, Ames, IA 50011, USA}
	\affiliation{Department of Physics and Astronomy, Iowa State University, Ames, IA 50011, USA}
	\author{R. Nath}
	\email{rnath@iisertvm.ac.in}
	\affiliation{School of Physics, Indian Institute of Science Education and Research, Thiruvananthapuram 695551, India}
	
	\begin{abstract}
		The structural and magnetic properties of the two-dimensional spin-$1/2$ depleted-kagome compound Cu$_7$(TeO$_3$)$_2$(SO$_4$)$_2$(OH)$_6$ are investigated using x-ray diffraction, magnetization, heat capacity, and $^1$H Nuclear Magnetic Resonance (NMR) measurements. From the analysis of magnetic susceptibility, we found a large Curie-Weiss temperature [$\theta_{\rm CW} = -50(2)$~K] and the co-existence of antiferromagnetic and ferromagnetic interactions. The value of $\theta_{\rm CW}$ gives an estimate of the average nearest-neighbour antiferromagnetic interaction of $J/k_{\rm B} \simeq 66$~K. The NMR relaxation rates ($1/T_1$ and $1/T_2$) exhibit a peak, providing evidence for a magnetic long-range order at $T^*\simeq 4$~K which appears to be canted antiferromagnetic type. Heat capacity also features a broad maximum at $T^*$ that moves towards higher temperatures with increasing magnetic field, reflecting defect induced Schottky anomaly. The frustration parameter $f_r = \lvert \theta_{\rm CW} \lvert/{T^{*}}\simeq 12.5$ renders the compound a highly frustrated low-dimensional magnet.
		%The frustrated nature of the compound is further deduced from the power-law behavior of heat capacity at low temperatures with an unusually \textbf{low exponent value ($\alpha \simeq 1.22$)} and persistent antiferromagnetic correlations with $\vec{q} \neq 0$ at high temperatures seen from the $^1$H spin-lattice relaxation rate $1/T_1(T)$.
	\end{abstract}
	\maketitle
	
	\section{Introduction}
	Exotic quantum phases thrive in materials that are low-dimensional and magnetically frustrated~\cite{Ramirez453,*Diep2013}. Frustration leads to degenerate states which has direct bearing on the ground state properties. For instance, it is expected to oust the conventional magnetic long-range-order (LRO) and results in diverse ground states, encompassing quantum spin-liquid (QSL), spin ice, spin glass etc~\cite{Savary016502}.
	%The degeneracy and ordering suppression result from the intrinsic incapacity of such materials to satisfy all spin-to-spin interactions. Sometimes, it can leave residual entropy down to absolute zero, causing a state of no-long-range ordering. Several unusual ground states, such as quantum spin liquid, spin ice, and spin glasses, can be found in such materials~\cite{Balents7286}.
	Antiferromagnets (AFM) with triangular motifs of magnetic ions as the basic building blocks constitute a class of geometrically frustrated magnets. These lattices range from normal edge-sharing triangular~\cite{Lal014429,Sebastian104425} to kagome~\cite{Zorko147201}, maple leaf~\cite{Schmalfu224405,*Haraguchi174439}, trillium~\cite{Yao146701,Kolay224405}, and pyrochlore geometries~\cite{Zhou227204}. 
	Among the geometrically frustrated magnets, the spin-1/2 kagome AFM composed of corner-sharing triangles are the prime candidates to host the exotic states of matter in two dimension. Further, quantum effects due to low dimensionality and low coordination number add up to the effect of magnetic frustration, leading the spin fluctuations to persist down to absolute zero temperature, a footprint of QSL~\cite{Clark207208}. %A few experimental examples are kapellasite [Cu$_3$Zn(OH)$_6$Cl$_2$]~\cite{Fak037208}, [NH$_4$]$_2$[C$_4$H$_{14}$N][V$_7$O$_6$F$_{18}$]~\cite{Clark207208}, and the celebrated candidate Herberthsmithite [ZnCu$_3$(OH)$_6$Cl$_2$]~\cite{Han406}.
	%In addition, compounds like (CH$_3$NH$_3$)$_2$KTi$_3$F$_{12}$~\cite{Jiang207}, Tb$_3$Ru$_4$Al$_{12}$~\cite{Upadhyay325601} show spin glass behavior, as well as Li$_9$Fe$_3$(P$_2$O$_7$)$_3$(PO$_4$)$_2$~\cite{Kermarrec157202} showing classical spin liquid behavior.
	Strikingly, many variants of the kagome lattice with slight structural modification have also been investigated owing to their frustrated nature e.g. hyper-kagome~\cite{Okamoto137207}, capped-kagome~\cite{Mohanty104424,Guchhait174447}, staircase-kagome~\cite{Morosan144403}, octa-kagome~\cite{Tang14057}, sphere-kagome~\cite{Rousochatzakis094420}, strip-kagome~\cite{Jeschke140410}, square-kagome~\cite{Fujihala3429}, tripod kagome~\cite{Dun157201,*Dum031069}, and stagome~\cite{Perez2173}. Some of them are proposed to host proximate QSL and other fascinating quantum phases at low temperatures~\cite{Okamoto137207,Fujihala3429,Penc197203,Magar054076}.
	%For example, hyperkagome iridate compounds like Na$_4$Ir$_3$O$_8$ and Li$_3$Ir$_3$O$_8$ show spin liquid behavior and semimetallic nature, respectively~\cite{Okamoto137207,Takayama075002}. Similarly, the square-kagome compound KCu$_6$AlBiO$_4$(SO$_4$)$_5$Cl shows properties of a gapless spin liquid~\cite{Fujihala3429}.
	The degree of structural distortion, site disorder, and defects are also the key ingredients that influence the ground state properties immensely~\cite{Paddison117,*Savary087203}.
	
	The kagome lattice compounds with depleted sites are almost an unexplored territory in the realm of geometrically frustrated magnets, as only a few compounds have been studied till date~\cite{Nishimoto196401,Ferhat155141}. For example, the $5/6$-depleted kagome lattice compounds Li$_2$InMo$_3$O$_8$ and Li$_2$ScMo$_3$O$_8$ show magnetic LRO at $T_{\rm N} \simeq 12$~K and QSL-like behavior, respectively~\cite{Haraguchi014409,Iida1826}. Similarly, LiZn$_2$Mo$_3$O$_8$ which can be considered as a $2/3$-depleted kagome lattice shows signatures of resonating valence-bond state at low temperatures~\cite{Sheckelton493,Mourigal027202}. More recently, magnetic properties of BaCu$_2$(PO$_4$)$_2$.(H$_2$O), which has a $1/3$ depleted kagome geometry, have been studied using thermodynamic and Nuclear Magnetic Resonance (NMR) techniques. It shows an incommensurate magnetic ordering at around $T_{\rm N} \simeq 10.5$~K~\cite{Singh125112}.
	
	\begin{figure*}
		\includegraphics[scale=0.125]{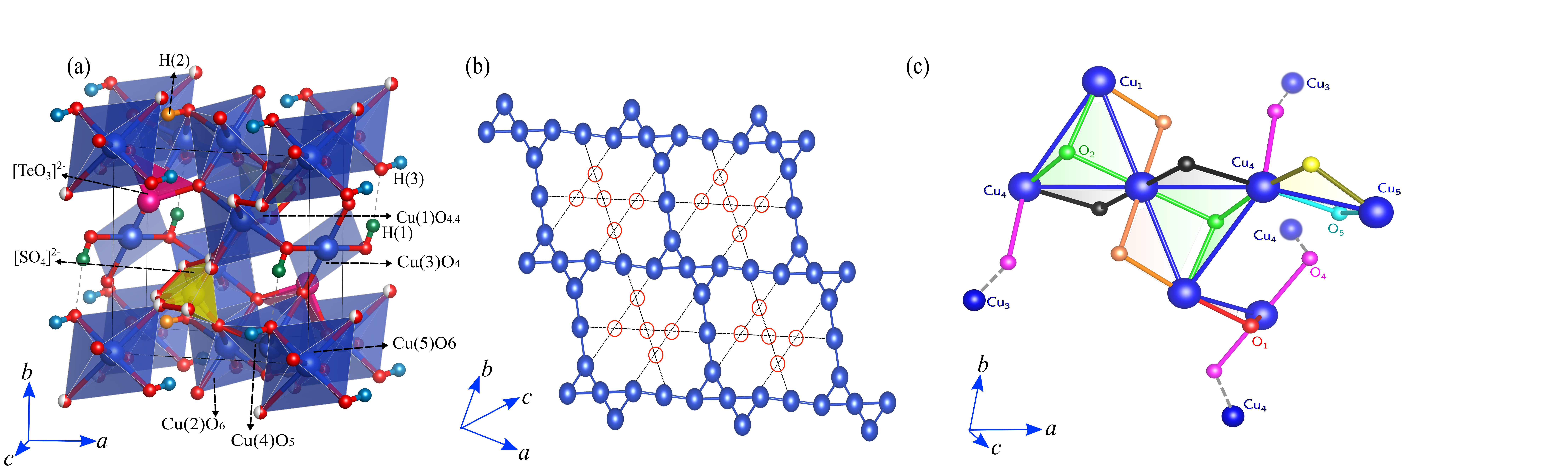}
		\caption{(a) Three-dimensional view of the crystal structure of CTSOH featuring Cu$^{2+}$ polyhedras (blue) and other atoms. (b) A layer of the depleted-kagome lattice formed by Cu$^{2+}$ ions. The hollow circles represent the depleted sites. (c) The smallest repeating unit in the depleted kagome layer, highlighting the interaction pathways. The corresponding bond angles and bond distances are tabulated in Table~\ref{TableI}.}
		\label{Fig1}
	\end{figure*}
	In this paper, we report the ground-state properties of a two-dimensional (2D) depleted $S=1/2$ kagome lattice compound Cu$_7$(TeO$_3$)$_2$(SO$_4$)$_2$(OH)$_6$ (abbreviated as CTSOH). The compound crystallizes in a triclinic space group $P{\bar 1}$ (No.~2). In the crystal structure shown in Fig.~\ref{Fig1}(a), there are five distinct Cu$^{2+}$ sites, coordinated with O atoms forming different polyhedra. Cu(1) constitutes a distorted Cu(1)O$_{4.45}$ octahedra with three partially and three fully occupied O atoms, Cu(2) forms a distorted Cu(2)O$_6$ octahedra, Cu(3) forms a regular Cu(3)O$_4$ square plaquette, Cu(4) forms a regular Cu(4)O$_5$ square pyramid, and Cu(5) makes a distorted Cu(5)O$_6$ polyhedra with four partially and four fully occupied O atoms.
	%\textbf{Cu(1) constitutes a highly distorted Cu(1)O$_{4.45}$ octahedra with three partially occupied O atoms (one-0.551~\&~two-0.449 occupancy), Cu(2) forms a distorted Cu(2)O$_6$ octahedra, Cu(3) forms a regular Cu(3)O$_4$ square, Cu(4) forms a regular Cu(4)O$_5$ square pyramid, and Cu(5) makes a distorted Cu(5)O$_6$ polyhedra with four partially occupied O atoms (two-0.551~\&~two-0.449 occupancy).}
	The polyhedra Cu(1)O$_{4.45}$, Cu(2)O$_6$, Cu(4)O$_5$, and Cu(5)O$_6$ are edge-shared sequentially while Cu(3)O$_4$ is corner-shared with Cu(1)O$_{4.45}$ and Cu(4)O$_5$. This entire arrangement constitutes kagome layers of Cu$^{2+}$ ions with 5/12 missing sites. Further, the corner sharing of Cu(3)O$_4$ and Cu(4)O$_5$ couples the kagome layers, providing inter-layer interaction. It is also to be noted that the partially occupied O atoms may possibly induce disorder/defects in the spin-lattice. The Cu-Cu bond distances and Cu-O-Cu bond angles within and between the kagome planes are listed in Table~\ref{TableI}. In addition, it also has [TeO$_{3}]^{2-}$, [SO$_{4}]^{2-}$, and H atoms, sitting at the interstitial space. Additional interaction paths among Cu$^{2+}$ ions within and between the kagome planes are mediated through TeO$_3$ and SO$_4$ units, respectively. A magnetic layer of the depleted kagome lattice formed by Cu$^{2+}$ ions is shown in Fig.~\ref{Fig1}(b), where 5/12 of the magnetic sites are absent as compared to a regular kagome lattice. The missing sites are marked by hollow circles. The smallest repeating unit of magnetic ions in the depleted kagome layer is shown in Fig.~\ref{Fig1}(c).
	%where corner shared triangular pairs are formed by two Cu(1), Cu(4), and one Cu(2) atom with Cu(2) as the common vertex. These triangular pairs are connected to neighboring pairs through Cu(5) atoms horizontally and Cu(3) atoms vertically.
	\begin{table}[h]
		\label{TableI}
		\caption{Details of Cu-O-Cu super-exchange pathways within and between the kagome layers in CTSOH.}
		\begin{tabular}{c c c }
			\hline\hline
			Bond &  Cu - Cu bond length  &  Bond angle \\ 
			&   (\AA)  &  ($^{\circ}$) \\ 
			\hline
			~ & Intralayer & ~ \\
			\hline
			Cu(1)-O(2)-Cu(2) & 2.93  & 94.9 \\
			Cu(1)-O(6)-Cu(2) & 2.93  & 97.5 \\ 
			Cu(1)-O(1)-Cu(3) & 3.19  & 113.8 \\ 
			Cu(2)-O(2)-Cu(4) & 3.10  & 103.8 \\ 
			Cu(2)-O(7)-Cu(4) & 3.10  & 88.8 \\
			Cu(4)-O(5)-Cu(5) & 2.86  & 89.5 \\
			Cu(4)-O(3)-Cu(5) & 2.86  & 95.5 \\
			\hline
			~ & Interlayer & ~ \\
			\hline
			Cu(3)-O(4)-Cu(4) & 3.85  & 128.8 \\
			\hline\hline
		\end{tabular}
	\end{table}
	
	The earlier preliminary magnetic measurements reported the absence of magnetic LRO down to 2~K, strong magnetic frustration, and a possible QSL candidate~\cite{Guo1830}. Herein, we carried out a detailed structural and magnetic studies of CTSOH using thermodynamic as well as local NMR probes. The magnetic measurements imply co-existing AFM and ferromagnetic (FM) interactions, a possible magnetic ordering at $T^* \simeq 4$~K, and strong magnetic frustration. The magnetic ordering appears to be canted AFM-type.  %Interestingly, heat capacity data suggest a disordered type state below $T^*$ in contrast to the peak observed in the NMR relaxation rates.
	%\textbf{While the compound is found to be a highly frustrated magnet, the NMR relaxation rates and NMR spectra point towards a partial ordering of spins at $T^* \simeq 4$~K. A significant fraction of spins still remains in the disordered or paramagnetic state at low temperatures. The unusual power-law character of the spin-lattice relaxation rate below $T^*$ also implies the persistence of sizeable spin fluctuations in the partially ordered state.}
	
	\section{Experimental details}
	\begin{figure}
		\includegraphics[width=\linewidth]{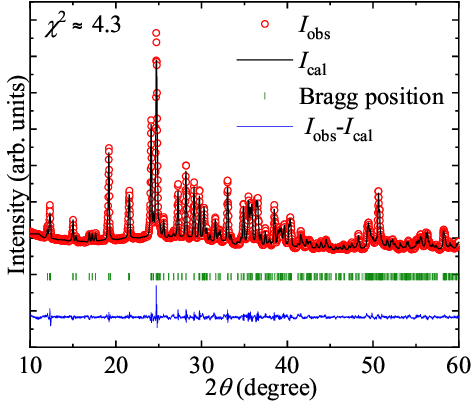}
		\caption{Room temperature powder XRD pattern of  CTSOH. The open circles represent the experimental data and the black solid line is the Rietveld fit. Bragg peaks are shown as vertical bars and the difference between the experimental and Rietveld fit is shown as a solid line at the bottom.}
		\label{Fig2}
	\end{figure}
	A polycrystalline sample of CTSOH was synthesized using hydrothermal synthesis route. A solution of an equimolar mixture of precursors was prepared using 0.239~g of CuSO$_4$, 0.076~g of K$_2$TeO$_3$, and 10~ml of deionized water in a 23~ml teflon jar. The teflon jar placed in a tightly closed autoclave was heated at 210$^{\circ}$C for 5~days and then slowly cooled for 3~days. Dark green polycrystalline clumps were sorted out with the help of a microscope and a tweezer which were subsequently ground into powder. The phase purity of the powder sample was verified by x-ray diffraction (XRD) performed using a PANalytical powder diffractometer (Cu\textit{K$_{\alpha}$} radiation, $\lambda_{\rm avg}\simeq 1.5418$~\AA). The acquired data are shown in Fig.~\ref{Fig2}. Rietveld refinement was performed using the \verb"FullProf" software package~\cite{RODRIGUEZ55}, for which the initial parameters were taken from the previously report~\cite{Guo1830}. The refined lattice parameters $a = 7.3830(3)$~\AA, $b = 7.6294(3)$~\AA, $c = 7.6520(4)$~\AA~, $\alpha = 75.18(2)^{\circ}$, $\beta = 75.89(2)^{\circ}$, and $\gamma = 84.17(2)^{\circ}$ agree well with the reported data~\cite{Guo1830}. This confirms that the compound stabilizes in a triclinic structure and the powder sample is single phase in nature.
	
	The DC magnetization data were measured by varying both temperature (1.8~K~$\leq T \leq 320$~K) and magnetic field (0~T~$\leq \mu_0H \leq 7$~T) using a superconducting quantum interference device (SQUID) magnetometer (Quantum Design, MPMS-3). Measurements below 1.8~K (down to 0.4~K) were performed using an additional $^3$He attachment to the SQUID. Heat capacity was measured on a $\sim 1$~mg sintered pellet in the physical property measurement system (PPMS, Quantum Design) employing the relaxation technique as a function of temperature as well as magnetic field. For measurements below 2~K (down to 0.4$~K$), a $^3$He insert was used. AC magnetization data were collected at different frequencies from 100~Hz to 10~kHz in	 an excitation field of $H_{\rm ac} = 10$~Oe using the ACMS option of PPMS.
	
	Nuclear magnetic resonance (NMR) measurements were performed using a phase-coherent spin-echo pulse spectometer on the $^{1}$H nuclei ($I = \frac{1}{2}$, gyromagnetic ratio $\gamma_{\rm N}/2\pi = 42.5774$~MHz/T). The NMR spectra were obtained by sweeping the magnetic field, keeping the frequency constant.  The $^1$H spin-lattice relaxation rate ($1/T_{1}$) was measured using the standard saturation recovery method. Similarly, the  $^1$H spin-spin relaxation rate ($1/T_{2}$) was measured by monitoring the decay of the echo integral with the variable spacing between the $\pi/2$ and $\pi$ pulses~\cite{Fukushima2018}.
	%The temperature-dependent NMR shift \textit{K(\rm T)} [=($H_{\rm ref}-H({\rm T}))/H({\rm T})$] was calculated by comparing the resonance field ($H$) of the sample with the resonance field ($H_{\rm ref}$) of the reference sample H$_2$O at each temperature.
	
	\section{Results and Discussion}
	\subsection{DC magnetization}
	\begin{figure}
		\includegraphics[width=\linewidth]{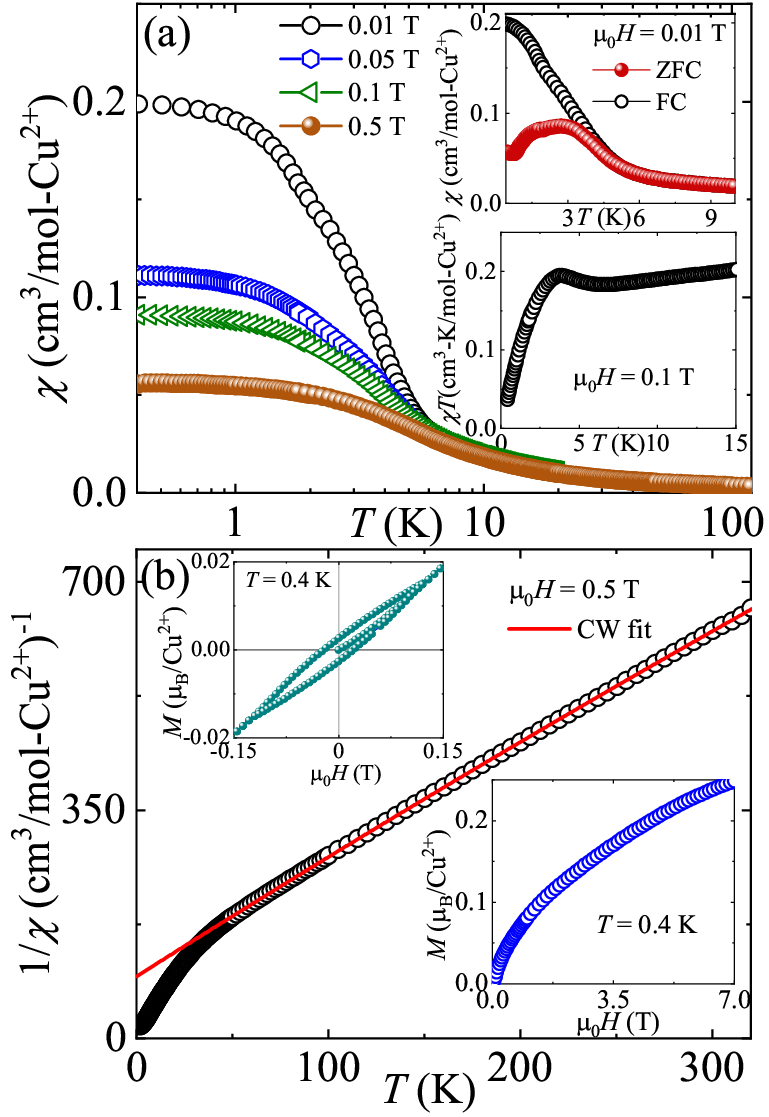}
		\caption{(a) $\chi$ vs $T$ in different fields. Upper inset: $\chi(T)$ measured in ZFC and FC conditions in a small field of $\mu_0H = 0.01$~T. Lower inset: $\chi T$ vs $T$ for $\mu_0 H = 0.1$~T.  (b) $1/\chi$ vs $T$ for $\mu_0 H = 0.5$~T. The solid line is the Curie Weiss fit. Upper inset: A complete $M$ vs $H$ isotherm at $T = 0.4$~K in the low field regime. Lower inset: Magnetic isotherm at $T = 0.4$~K from 0 to 7~T.}
		\label{Fig3}
	\end{figure}
	Temperature-dependent DC magnetic susceptibility $\chi$ measured in different applied fields is shown in Fig.~\ref{Fig3}(a). In a small field, $\chi(T)$ exhibits an upward trend below $T^* \simeq 4$~K which apparently saturates at low temperatures. As the magnetic field is increased, a strong suppression of $\chi(T)$ is observed below $T^*$. These features do not provide evidence for a conventional AFM LRO, rather, it is a signature of the onset of FM correlations. $\chi(T)$ measured under zero-field cooled (ZFC) and field-cooled (FC) protocols show a clear splitting at $T^* \simeq 4$~K for $\mu_0H = 0.01$~T [upper inset of Fig.~\ref{Fig3}(a)].
	
	As presented in Fig.~\ref{Fig3}(b), the inverse susceptibility $1/\chi$ for $\mu_0 H = 0.5$~T is linear at high temperatures. $1/\chi$ in the linear regime ($T > 190$~K) is fitted using the modified Curie-Weiss (CW) law
	\begin{equation}
		\chi(T)=\chi_0+\frac{C}{(T-\theta_{\rm CW})},
		\label{Eq1}
	\end{equation}
	where $\chi_0$ is the temperature-independent susceptibility, defined by $\chi_0 = \chi_{\rm core} + \chi_{\rm VV}$. Here, $\chi_{\rm core}<0$ is the core diamagnetic and $\chi_{\rm vv}>0$ is the Van Vleck paramagnetic contributions present in the compound. $C$ is the Curie constant and $\theta_{\rm CW}$ is the characteristic CW temperature.
	The fit results in $\chi_0 = 1.0(2) \times 10^{-4}$~cm$^3$/mol-Cu$^{2+}$, $C = 0.51(3)$~cm$^3$K/mol-Cu$^{2+}$, and $\theta_{\rm CW} =-50(2)$~K. Adding the core diamagnetism of individual ions in the formula unit Cu$_7$(TeO$_3$)$_2$(SO$_4$)$_2$(OH)$_6$, we obtained $\chi_{\rm core} = -4\times 10^{-4}$~cm$^3$/mol~\cite{Bain532}. Next, by subtracting $\chi_{\rm core}$ from $\chi_0$, we got $\chi_{\rm VV} \simeq 5 \times 10^{-4}~$cm$^3$/mol, which is comparable with other cuprate compounds~\cite{Motoyama3212,Nath174436}. From the value of $C$, the effective magnetic moment is calculated to be $\mu_{\rm eff} = \sqrt{\frac{3k{\rm_B}C}{N_{\rm A}}}\mu_{\rm B} = 2.0(3)~\mu_{\rm B}$ (where, $N_{\rm A}$ is the Avogadro's number and $k_{\rm B}$ is the Boltzmann constant). This is slightly higher than the expected value of $1.73~\mu_{\rm B}$ $[= g\sqrt{S(S+1)}]$ for a free $S=1/2$ and corresponds to $g = 2.32(4)$. Indeed, such a higher $g$-value is typically observed in most of the Cu$^{2+}$ based systems~\cite{Nath014407,Nath054409,Sebastian064413,Lebernegg174436}.
	
	The negative value of $\theta_{\rm CW}$ confirms the dominant AFM exchange coupling among the Cu$^{2+}$ ions. With a tentative assignment of $T^* \simeq 4$~K as the ordering temperature, the frustration ratio is calculated to be $f_r = \lvert \theta_{\rm CW} \lvert /{T^{*}}\simeq 12.5$. This characterizes the material as a highly frustrated magnet. In addition, the magnitude of $\theta_{\rm CW}$ represents the overall energy scale of the exchange couplings in the system as, $\lvert \theta_{\rm CW} \lvert = \frac{JzS(S+1)}{3k_{\rm B}}$~\cite{Domb296}.
	%\begin{equation}
	%    \lvert \theta_{\rm CW} \lvert = \frac{JzS(S+1)}{3k_B}.
	%    \label{Eq2}
	%\end{equation}
	Here, $J$ is the average strength of the nearest-neighbour exchange couplings and $z=3$ is the average number of nearest-neighbor spins coupled via exchange coupling. The corresponding Hamiltonian of the Heisenberg model is $H = J \sum_{<i,j>} \vec{S}_i \cdot \vec{S}_j$. From this mean-field expression, using the experimental value of $\theta_{\rm CW}$, we obtained $J/k_{\rm B} \simeq 66$~K.
	
	A magnetic isotherm measured at $T = 0.4$~K is shown in the inset of Fig.~\ref{Fig3}(b). Around the zero-field, it evinces a small hysteresis as expected, since ZFC-FC $\chi(T)$ shows a bifurcation. As the field increases, $M$ shows a weak bend and then increases linearly with $H$. This linear increase and the absence of saturation even at 7~T point towards the dominant AFM interactions. Similarly, the initial bend can be ascribed to the saturation of a weak FM interaction present in the compound. Further, the tiny hysteresis and lack of saturation also rule out a FM ordering and point towards a canted-AFM type ordering at $T^*$ that induces a weak ferromagnetism.
	
	In order to find further evidence for the coexistence of FM and AFM interactions, we plotted $\chi T$ vs $T$ in the lower inset of Fig.~\ref{Fig3}(a). With decreasing temperature, $\chi T$ decreases continuously at high temperatures. Below 10~K, it shows an increase, passes through a broad maximum at around 3~K, and then falls gradually toward zero. These features in the $\chi T$ vs $T$ plot are a clear signature of the dominance of FM and AFM correlations at high and low-temperatures regimes, respectively, separated by the broad maximum~\cite{Savina104447,Mohanty134401}. Since the Cu$^{2+}$ - Cu$^{2+}$ interactions involve superexchange via oxygen atoms, one can analyze it in terms of the Goodenough-Kanamori-Anderson rules~\cite{Kanamori87,*Goodenough564}. According to which it favours AFM interaction for $\angle \text {Cu-O-Cu} \simeq 95 - 180^{\degree}$ and FM interaction for $\leq 95^{\degree}$. A close inspection of the bond angles (see Table~\ref{TableI}) divulges that there is a difference in the bond angles and one would anticipate the intra-layer interactions Cu(2)-Cu(4) and Cu(4)-Cu(5) to be FM while the other ones are AFM in nature. However, a precise knowledge about the nature of the exchanges requires the estimation of individual exchange couplings using density functional theory calculations.
	%A broad crossover at TCros separates two different magnetic regimes with predominantly ferro- and antiferromagnetic correlations above and below this temperature. 
	%Thus, the coexistence of FM and AFM interactions is inferred from the small negative value of $\theta_{\rm CW}$ as well as from the $\chiT$ behavior.
	%Moreover, despite a clear splitting between ZFC and FC $\chi(T)$, the ac susceptibility does not show any significant frequency dependence (not shown). This rules out a spin-glass type phase. Hence, the observed splitting could be due to a small amount of defects present in the sample~\cite{Dey140405,Okamoto137207}.
	%originating from that only a small fraction of spins are in the disordered state, resulting in ZFC and FC splitting in $\chi(T)$~\cite{Dey140405,Okamoto137207}.
	
	\subsection{AC susceptibility}
	\begin{figure}
		\includegraphics[width=\linewidth]{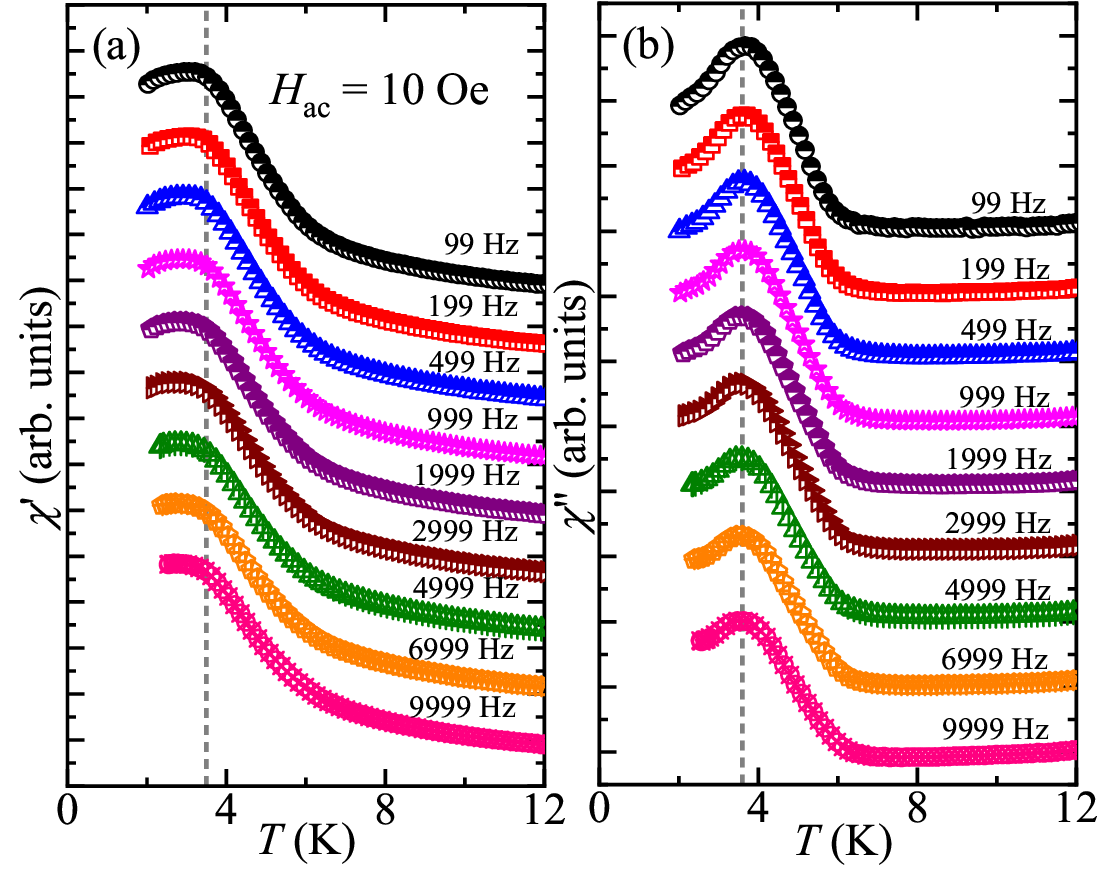}
		\caption{(a) Real part ($\chi^{\prime}$) and (b) imaginary part ($\chi^{\prime\prime}$) of the AC susceptibility as a function of $T$, measured at different frequencies from 99~Hz to 9999~Hz. The data sets in different frequencies are vertically offset for clarity.}
		\label{Fig4}
	\end{figure}
	The observed bifurcation of DC $\chi(T)$ measured under ZFC and FC conditions and small hysteresis in the magnetic isotherm below $T^*$ suggest three possible scenarios: (i) spin-glass transition or (ii) ferrimagnetic order or (iii) canted AFM order. To access the spin-glass/freezing, we measured AC susceptibility in different frequencies. As shown in Fig.~\ref{Fig4}, both real $\chi^{\prime}(T)$ and imaginary $\chi^{\prime\prime}(T)$ parts of AC susceptibility feature a broad peak near $T^* \simeq 3.8$~K that coincides with the bifurcation point of ZFC-FC $\chi(T)$. Surprisingly, this peak in both $\chi^{\prime}(T)$ and $\chi^{\prime\prime}(T)$ is found to be frequency independent, clearly ruling out a spin-glass scenario. Furthermore, $1/\chi$ vs $T$ in Fig.~\ref{Fig3}(b) exhibits a perfect linear behaviour (instead of a negative curvature) at high temperatures, excluding the possibility of a ferrimagnetic order~\cite{Nath224513}. Thus, all these observations suggest the onset of a canted AFM ordering at $T^*$, possibly driven by anisotropic Dzyaloshinskii-Moriya interaction~\cite{Kumar014410,Gaultois186004}. It is to be noted that presence of disorder/defect can also lead to the splitting of ZFC and FC $\chi(T)$~\cite{Dey140405,Okamoto137207}.
	%However, the possibility of defects cannot be overruled since they also lead to the splitting of ZFC and FC $\chi(T)$ and a weak magnetic hysteresis~\cite{Dey140405,Okamoto137207}.}

\subsection{Heat capacity}
\begin{figure}
	\includegraphics[width=\linewidth]{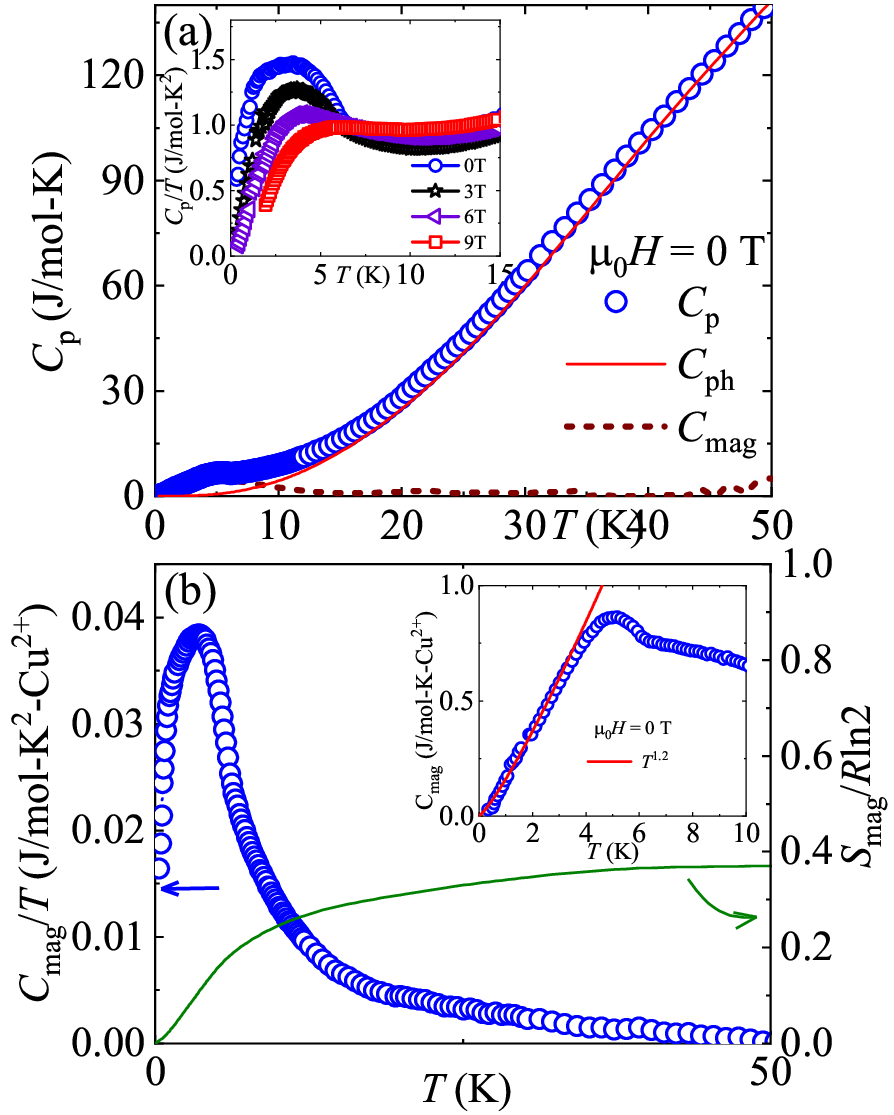}
	\caption{(a) Zero-field $C_{\rm p}$ vs $T$ along with the Debye-Einstein fit (solid line). The dashed line represents the magnetic heat capacity ($C_{\rm mag}$). Inset: $C_{\rm p}(T)$ around the low temperature anomaly, in different magnetic fields. (b) $C_{\rm mag}/T$ and normalized magnetic entropy ($S_{\rm mag}$/Rln2) vs $T$ on the left and right $y$-axes, respectively. Inset: Zero-field $C_{\rm mag}$ vs $T$ in the low temperature regime and the solid line is a power-law fit.}
	\label{Fig5}
\end{figure}
Temperature-dependent heat capacity [$C_{\rm p}(T)$] measured in zero-field is presented in Fig.~\ref{Fig5}(a). At high temperatures, it is dominated by the phonon excitations while at low temperatures, it is mostly the magnetic contribution. As seen from the inset of Fig.~\ref{Fig5}(a), the zero-field $C_{\rm p}/T$ shows a broad hump with center of gravity at $T^* \simeq 3.8$~K in contrast to a $\lambda$-type anomaly, typically expected for an AFM LRO.
	%Thus, the absence of a $\lambda$-type anomaly rules out a conventional magnetic LRO.
	%Interestingly, this $T^*$ coincides with the bifurcation point of the ZFC and FC $\chi(T)$ which possibly indicates the development of short range correlations.
Such as broad maximum in zero-field heat capacity is also indicative of an AFM short range order, commonly seen in highly frustrated magnets at low temperatures. Few representative examples are NiGa$_2$S$_4$~\cite{Nakatsuji1697}, Na$_4$Ir$_3$O$_8$~\cite{Okamoto137207}, Ba$_3$CuSb$_2$O$_9$~\cite{Zhou147204}, and Sc$_2$Ga$_2$CuO$_7$~\cite{Kumar180411}. Further, the position of the broad maximum is found to shift towards higher temperatures with magnetic field, which could suggest a Schottky type effect possibly due to a small fraction of disordered Cu$^{2+}$ spins and/or defects~\cite{Kumar180411}.

To separate the magnetic contribution from the total heat capacity, the phononic contribution [$C_{\rm ph}(T)$] is estimated using one Debye and four Einstein terms corresponding to one heavy and four lighter elements in the formula unit as~\cite{Gopal2012,Lal014429}
\begin{equation}
	C_{\rm ph}(T)=f_{\rm D}C_{\rm D}(\theta_{\rm D},T)+\sum_{i = 1}^{4}g_{i}C_{{\rm E}_i}(\theta_{{\rm E}_i},T).
	\label{Eq3}
\end{equation}
The first term in Eq.~\eqref{Eq3} is the Debye model that accounts for the acoustic mode has the form
\begin{equation}
	C_{\rm D} (\theta_{\rm D}, T)=9nR\left(\frac{T}{\theta_{\rm D}}\right)^{3} \int_0^{\frac{\theta_{\rm D}}{T}}\frac{x^4e^x}{(e^x-1)^2} dx.
	\label{Eq4}
\end{equation}
Here, $x=\frac{\hbar\omega}{k_{\rm B}T}$ with $\omega$ being the frequency of oscillation, $R$ is the real gas constant, and $\theta_{\rm D}$ is the Debye temperature. Likewise, the second term is the Einstein term, which accounts for the optical modes and can be written as
\begin{equation}
	C_{\rm E}(\theta_{\rm E}, T) = 3nR\left(\frac{\theta_{\rm E}}{T}\right)^2 
	\frac{e^{\left(\frac{\theta_{\rm E}}{T}\right)}}{\left[e^{\left(\frac{\theta_{\rm E}}{T}\right)}-1\right]^{2}}.
	\label{Eq5} 
\end{equation}
Here, $\theta_{\rm E}$ is the characteristic Einstein temperature. The coefficients $f_{\rm D}$, $g_1$, $g_2$, $g_3$, and $g_4$ represent the weight factor of the respective terms. Since there are 37 atoms in the formula unit and each atom has three degrees of freedom, it should have a total of 111 modes. Further, since there are three acoustic modes (or, Debye modes), we have fixed the value of Debye co-efficient ($f_{\rm D}$) to $3/111 \simeq 0.027$. Likewise, we have 108 optical modes, hence, we chose the coefficients $g_1 \simeq 0.06$, $g_2 \simeq 0.24$, $g_3 \simeq 0.48$, and $g_4 \simeq 0.193$ such that, their sum is equal to 108/111. The obtained characteristic temperature scales are $\theta_{\rm D} = 75(2)$~K, $\theta_{\rm E1} = 100(3)$~K, $\theta_{\rm E2} = 180(8)$~K, $\theta_{\rm E3} = 600(10)$~K, and $\theta_{\rm E4} = 2000(100)$~K. The simulated $C_{\rm ph}$ (solid line) in Fig.~\ref{Fig5}(a) matches well with the experimental $C_{\rm p}(T)$ in the high temperature regime. The obtained $C_{\rm mag}(T)$ after subtracting $C_{\rm ph}(T)$ from $C_{\rm p}(T)$ is presented as a dashed line in Fig.~\ref{Fig5}(a).

Figure~\ref{Fig5}(b) shows $C_{\rm mag}/T$ as a function of temperature which clearly demonstrates a very broad maximum at $T^*$. The magnetic entropy was calculated by integrating $C_{\rm mag}/T$ as $S_{\rm{mag}}(T) = \int_{\rm 0\,K}^{T}\frac{C_{\rm {mag}}(T')}{T'}dT'$. Surprisingly, the magnetic entropy at 50~K (in the saturated state) is about $\sim 2.13$~J/mol-K, which is only 37\% of the total expected value of $S_{\rm mag} = R \ln(2) = 5.76$~J/mol-K for a spin-$1/2$ system. This type of reduced entropy is typically observed in case of disordered systems, such as in QSL candidates~\cite{Zhou147204,Kumar180411}. 
%This is because, in the QSL systems, due to high degeneracy of the low-energy states, a large residual entropy still remains at low temperatures. 
The existence of partial disorder can also gives rise to this kind of unusually low entropy change~\cite{Takeda2017}.

%Magnetic heat capacity in a 3D AFM ordered state is expected to follow a power law ($C_{\rm mag} \propto T^\alpha$) behavior with an exponent $\alpha = 3$~\cite{Nath024431}. On the other hand, a reduced value of $\alpha < 3$ is envisaged in case of gapless QSLs~\cite{Ran1177205,Zhou147204,Kundu267202}.
%While Dirac SLs are expected to show a quadratic ($\alpha = 2$) temperature dependency, a more conventional fermionic spinon excitations lead to a temperature linear behaviour ($\alpha = 1$) at low temperatures~\cite{Ran1177205,Zhou147204,Kundu267202}.
In CTSOH, the zero-field $C_{\rm mag}$ below 2.5~K could be fitted by a power law of the form $C_{\rm mag} \propto T^\alpha$ with an exponent $\alpha = 1.2(2)$ and the coefficient $\gamma = 1022(1)$~mJ/mol-K. In a 3D AFM ordered state, one expects the value $\alpha = 3$~\cite{Nath024431}. On the other hand, a reduced value of $\alpha < 3$ is envisaged in case of gapless QSLs and other systems exhibiting unconventional spin dynamics~\cite{Clark207208,Kundu267202,Somesh104422}.
%Despite a magnetic LRO, such a significant reduction in the $\alpha$ value for CTSOH can possibly be attributed to the effect of disorder or defects. A precise knowledge on this issue requires further investigations on good quality single crystals.

%Though, this value of $\alpha$ differs appreciably from the classical AFM value, it is remarkably similar to the celebrated \textbf{possible} QSL candidates such as, herbertsmithite ($\alpha = 1.3$)~\cite{Vries157205} and Vanadium Oxyfluoride ($\alpha = 1.2$)~\cite{Clark207208}. Thus, the observed reduced value of $\alpha$ suggests an unconventional ground state in CTSOH. A reduced value of $\alpha$ is reported even in many frustrated magnets due to the existence of significant spin fluctuations in the ordered state~\cite{Koteswararao035141,Bonville7777}. Furthermore, $\gamma$ is related to the spinon density of states at the spinon Fermi surface and a large value of $\gamma$, similar to ours, has been reported for several potential QSL candidates~\cite{Cheng197204,Dey174411,BhattacharyaL060403}.}

\subsection{$^{1}$H NMR}
In order to obtain further information about the static and dynamic magnetic properties of CTSOH, we performed $^1$H NMR measurements. It is noted that the compound CTSOH contains three inequivalent H sites (H1, H2, and H3 shown in Fig. \ref{Fig1}). 
Each $^1$H site couples to two neighboring Cu$^{2+}$ ions, either through Cu-O-H–O–Cu or Cu–O–H–O–H–O-Cu pathways. 

\subsubsection{$^{1}$H NMR spectra above $T^*$}
\begin{figure*}
\includegraphics[scale=0.3]{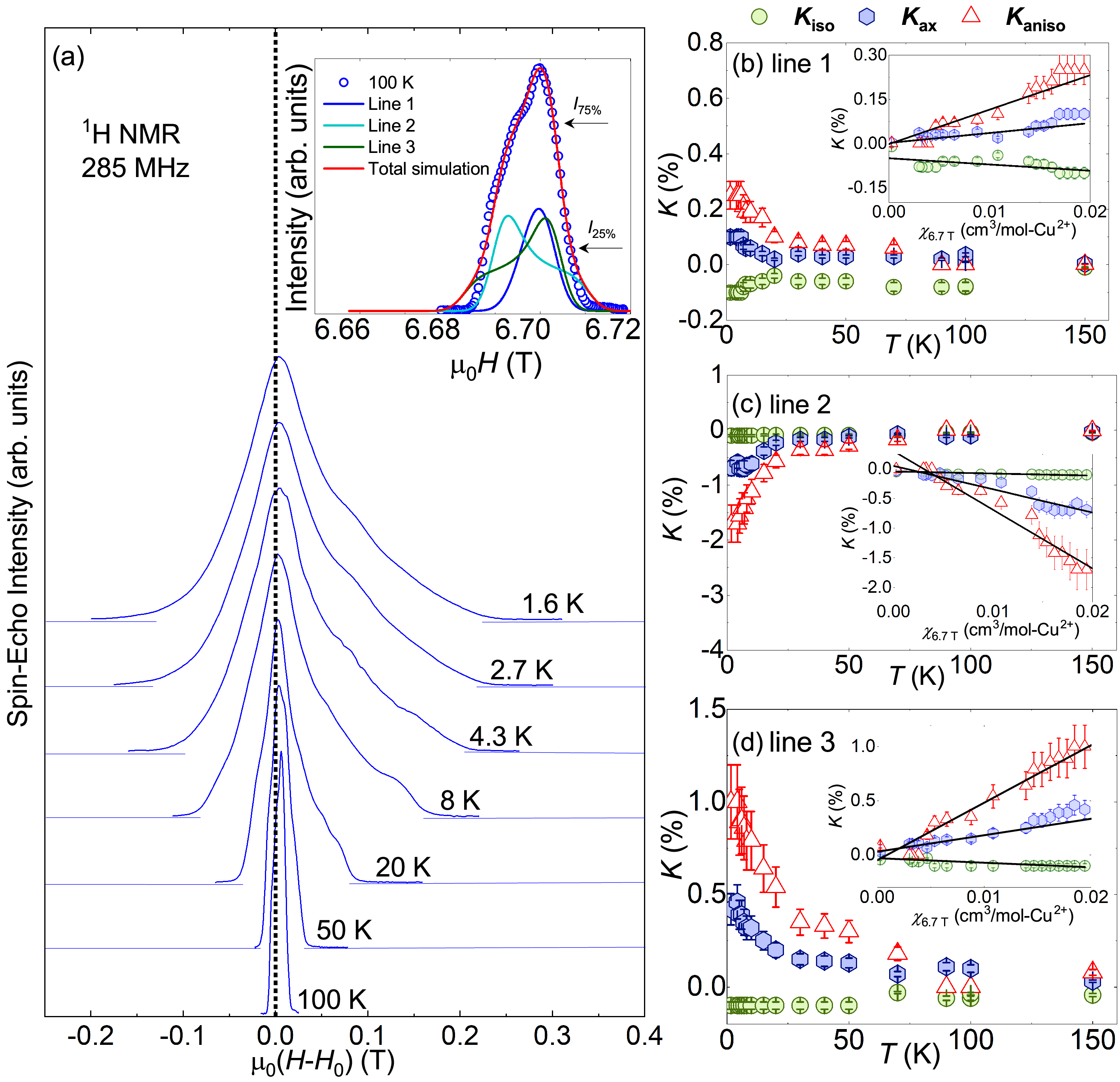}
\caption{(a) Temperature evolution of $^{1}$H NMR spectra measured in a frequency of $285$~MHz down to $T = 1.6$~K. The vertical dashed line corresponds to the zero-shift position ($H_0 = 6.694$~T). Inset: NMR spectra at $T=100$~K. The red curve is the sum of three calculated lines (line 1 in blue, line 2 in cyan, and line 3 in olive) and the arrows mark the positions where $1/T_1$ is measured. Temperature dependence of NMR shift ($K$) with the inset showing $K$ vs $\chi$ (measured at 6.7~T) plots for (b) Line 1, (c) Line 2, and (d) Line 3.}
\label{Fig6}
\end{figure*}
\begin{figure}
\includegraphics[width=\linewidth]{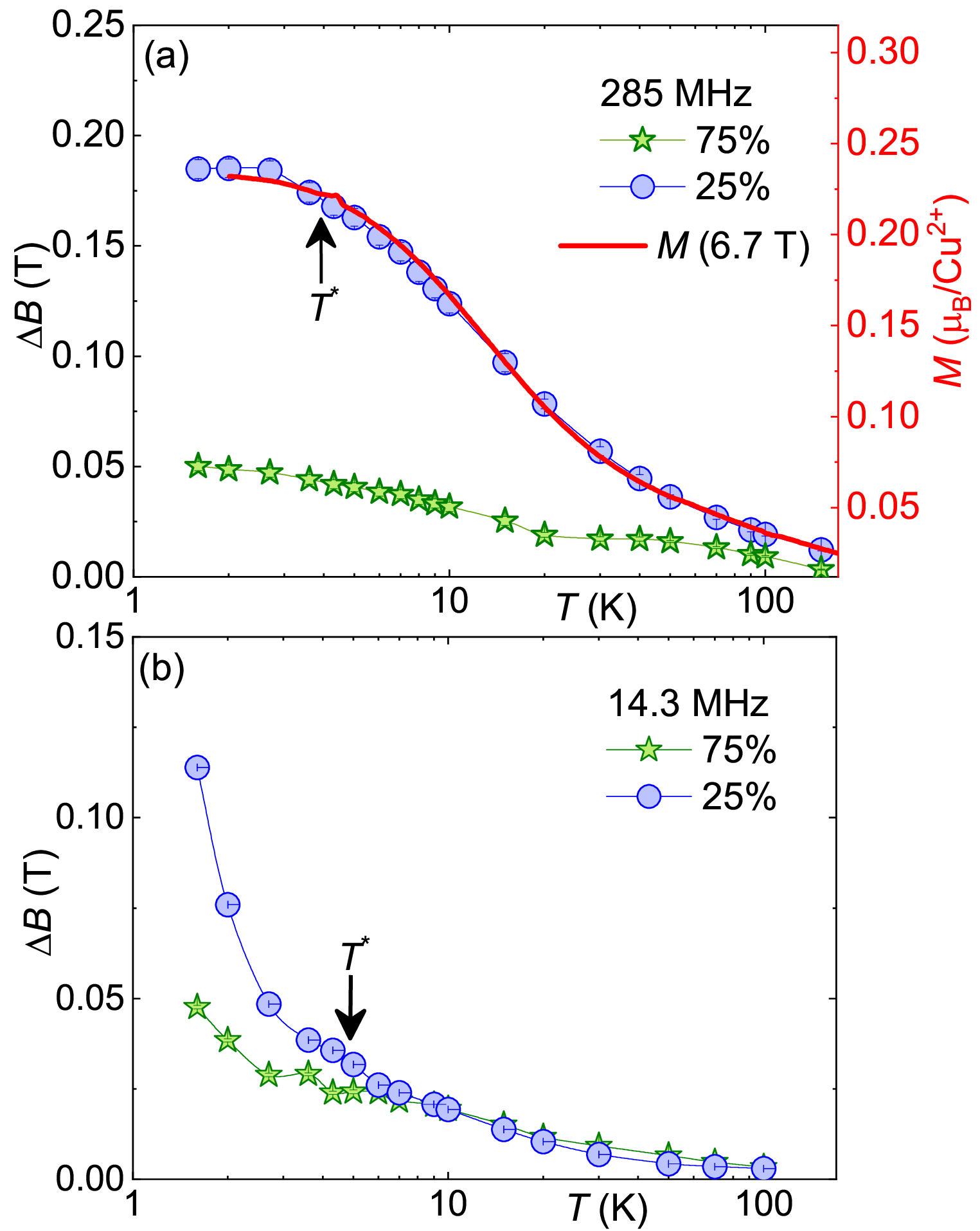}
\caption{(a) Temperature dependence of the linewidth ($\Delta B$) at 25$\%$ ($I_{25\%}$) and 75$\%$ ($I_{75\%}$) of the maximum intensity of the NMR spectra measured at $f = 285$~MHz.  Right $y$-axis shows the magnetization as a function of temperature. (b) Temperature dependence of the linewidth ($\Delta B$) measured at $f = 14.3$~MHz. The actual $\Delta B$ data at $I_{75\%}$ are multiplied by a factor of 2 for both the frequencies to improve the visibility.}
\label{Fig7}
\end{figure}
Figure~\ref{Fig6}(a) shows the typical $^{1}$H NMR spectra measured at various temperatures under a magnetic field of $\sim$6.69 T. At high temperatures, the spectra display a relatively narrow but asymmetric line shape with two shoulders along with the main peak. This feature can be attributed to three distinct $^1$H sites in the crystal structure as described below. As the temperature decreases, the spectra become broader.
%, and the broad spectra are found to exhibit not much change below $T^*$ $\sim 4$~K.

The observed NMR spectra were reproduced by the superposition of three anisotropic $^{1}$H NMR lines, as shown in the inset of Fig.~\ref{Fig6}(a). The anisotropic powder-patterns were calculated  with isotropic ($K_{\text{iso}}$), axial ($K_{\text{ax}}$), and anisotropic ($K_{\text{aniso}}$) components in  the NMR shift ($K$), where $K$ is described by \cite{Slichter2013}
\begin{equation}
K = K_{\text{iso}} + K_{\text{ax}}(3\cos^2\theta - 1) + K_{\text{aniso}}\sin^2\theta \cos 2\phi .
\label{eq:NMR}
\end{equation}
Here, $\theta$ and $\phi$ are the polar and azimuthal angles between the external magnetic field and the principal axis of the hyperfine field tensor at each H site, respectively. While the direction of the principal axis for each H site is unknown, this will not be an issue in calculating the powder-pattern spectrum. 
The blue (line 1), cyan (line 2), and olive (line 3) curves are the calculated spectra using different values of the set of parameters ($K_{\text{iso}}$, $K_{\text{ax}}$, $K_{\text{aniso}}$) = (-0.078\%, 0.036\%, $\sim$0), (-0.054\%, -0.1\%, $\sim$0), and (-0.036\%, 0.1\%, $\sim$0), respectively, with appropriate broadening for each line. 
The red curve which is the sum of three calculated lines with nearly equal intensity, roughly reproduces the observed spectral shape for $T > 6$~K without any additional component. At low temperatures, the line becomes broad and there are multiple features due to three inequivalent H-sites. Therefore, it was difficult to make any assessment about the defects, if at all present, from the spectral shape.
%beyond what is expected from the three inequivalent proton sites. This suggests that any such defects, if present, do not contribute significantly to the NMR spectra. 
Nevertheless, we cannot rule out a small concentration of local defects, and high-resolution techniques such as magic-angle spinning NMR or pair-distribution-function analysis would be required to identify them unambiguously.

From the fitting of the spectra, we determined the temperature dependencies of  $K_{\text{iso}}$, $K_{\text{ax}}$, and  $K_{\text{aniso}}$ which are shown in Figs.~\ref{Fig6}(b), (c), and (d) for line 1, line 2, and line 3, respectively. Although, the observed spectra down to 1.6~K were roughly reproduced by the simulations, it should be noted that, there are a large number of parameters (3 components of $K$ for each line: 9 parameters in total). Therefore, the resulting 
%we could not precisely determine the value of each parameter from the fitting, 
parameters are having relatively large uncertainties ($\sim$ 20 \%). Nevertheless, we estimated the corresponding hyperfine coupling constants for each $^1$H-NMR line using the relation
\begin{equation}
K = K_0 + \dfrac{\mathcal{A}_{\rm hf}}{N_{\rm A}\mu_{\rm B}}\chi,
\label{k-chi}
\end{equation}
where $K_0$ is the temperature-independent orbital contribution, $N_{\rm A}$ is Avogadro’s number, and $\mathcal{A}_{\rm hf}$ is the hyperfine coupling constant. We plotted $K$ as a function of $\chi$ for each NMR shift component ($K_{\text{iso}}$, $K_{\text{ax}}$, and $K_{\text{aniso}}$) of all the three lines in the insets of Figs.~\ref{Fig6}(b), (c), and (d). From a linear fit to the $K$–$\chi$ data using Eq.~\eqref{k-chi}, we obtained the isotropic ($\mathcal{A}_{\rm hf}^{\rm iso}$), axial ($\mathcal{A}_{\rm hf}^{\rm ax}$), and anisotropic ($\mathcal{A}_{\rm hf}^{\rm aniso}$) hyperfine coupling constants for three lines, which are summarized in Table~\ref{tab:hyperfine}.

\begin{table}[h]
\caption{\label{tab:hyperfine}
The estimated  isotropic ($\mathcal{A}_{\rm hf}^{\rm iso}$), axial ($\mathcal{A}_{\rm hf}^{\rm ax}$), and anisotropic ($\mathcal{A}_{\rm hf}^{\rm aniso}$) hyperfine coupling constants for three $^1$H lines in units of T/$\mu_{\rm B}$.}
\begin{ruledtabular}
\begin{tabular}{c c c c c c c}
	Line & $\mathcal{A}_{\rm hf}^{\rm iso}$ (T/$\mu_{\rm B}$) & $\mathcal{A}_{\rm hf}^{\rm ax}$ (T/$\mu_{\rm B}$)  &   $\mathcal{A}_{\rm hf}^{\rm aniso}$ (T/$\mu_{\rm B}$) \\
	\hline
	1 &  -0.017(3)    & 0.019(4)    &   0.06(1) \\
	2 &  -0.019(4) &   -0.22(4)      &  -0.5(1)   \\
	3 & -0.023(5)  & 0.08(1)   &  0.29(6)  \\
\end{tabular}
\end{ruledtabular}
\end{table}

\begin{table}[h]
\caption{\label{dipolar:coupling}
Calculated values of dipolar field for three $^1$H sites in units of T/$\mu_{\rm B}$.}
\begin{ruledtabular}
\begin{tabular}{cccr}
	& $\mathcal{A}_{\rm hf}^{\rm iso}$ (T/$\mu_{\rm B}$)  & $\mathcal{A}_{\rm hf}^{\rm ax}$ (T/$\mu_{\rm B}$)  &   $\mathcal{A}_{\rm hf}^{\rm aniso}$ (T/$\mu_{\rm B}$) \\
	\hline
	H2 &  0  &  0.020    &  0.048 \\
	H1 &  0  & -0.10   &  -0.25 \\
	H3 &  0  & 0.045     &  0.12 \\
\end{tabular}
\end{ruledtabular}
\end{table}

Since the isotropic parts of the hyperfine coupling constants are relatively small, we calculated the classical dipolar field at each  $^1$H site using lattice summation. The calculated values for the three $^1$H sites are summarized in Table~\ref{dipolar:coupling}. The sign and magnitude of $\mathcal{A}_{\rm hf}^{\rm ax}$ and $\mathcal{A}_{\rm hf}^{\rm aniso}$ allow us to assign line 2 to the H1 site, as both components are negative only for this site. Similarly, lines 1 and 3 can be assigned to H2 and H3, respectively. In the case of powder samples, classical dipolar fields cause line broadening (and possibly shoulders) but do not yield any net shift of the resonance line. Therefore, the finite values of $\mathcal{A}_{\rm hf}^{\rm iso}$ for all three H sites must originate from the transferred hyperfine field~\cite{Owen791}, likely due to the overlapping between H-1$s$ and neighboring Cu-3$d$ orbitals mediated through the O-2$p$ orbitals along the Cu–O–H–O–Cu and Cu–O–H–O–H–O–Cu paths. Furthermore, slightly different $\mathcal{A}_{\rm hf}^{\rm iso}$ values for three H-sites can be attributed to different extents of orbital overlaps.

Next, we focus on the temperature dependence of the spectral linewidth. 
To capture the broadening behavior of the anisotropic NMR line shape, we show the temperature dependence of the linewidth $\Delta B$ at 75\% and 25\% of the peak intensity in Fig.~\ref{Fig7}(a), together with the temperature dependence of magnetization ($M$) measured at 6.7~T. Although the absolute values of $\Delta B$ differ significantly between two positions, both exhibit a similar temperature dependence, closely replicating the behavior of magnetization, as expected. Moreover, $\Delta B$ does not show a clear increase below $T^*$ which is likely due to the effect of a strong magnetic field ($\mu_0H$ $\sim 6.69$~T). However, $\Delta B(T)$ measured in a lower field of $\mu_0H \sim 0.34$~T, exhibits a weak anomaly near $T^*$ and a small increase below that [Fig.~\ref{Fig7}(b)].
% In the paramagnetic regime, $\Delta B$ is generally expected to scale with the magnetization $M$. We observe a strong linear correlation between $\Delta B$ and $M$ across the entire temperature range for $\Delta B_{25\%}$ part, suggesting the presence of weak internal fields arising from the Cu$^{2+}$ spins. 

% To quantify this, we exploit the linear relation $\Delta B\propto A_{\rm hf}M$, and plot $\Delta B$ versus $M$ in Fig.~\ref{Fig6}(b), using temperature as an implicit parameter. From the slopes, we estimate the hyperfine coupling constants to be $A_{\rm hf}^{75\%}\simeq0.076(1)$~T/$\mu_{\rm B}$ and $A_{\rm hf}^{25\%}\simeq 0.407(1)$~T/$\mu_{\rm B}$. The obtained hyperfine coupling constants represent the maximum possible values expected for CTSOH. While the major contribution is consistent with dipolar interactions, the magnitudes suggest that a small but finite transferred hyperfine component may also contribute to the total hyperfine coupling, likely arising from weak orbital overlap between hydrogen and neighboring Cu$^{2+}$ ions.

\subsubsection{$^1$H NMR spectra below $T^*$}
\begin{figure}
\includegraphics[width=\linewidth]{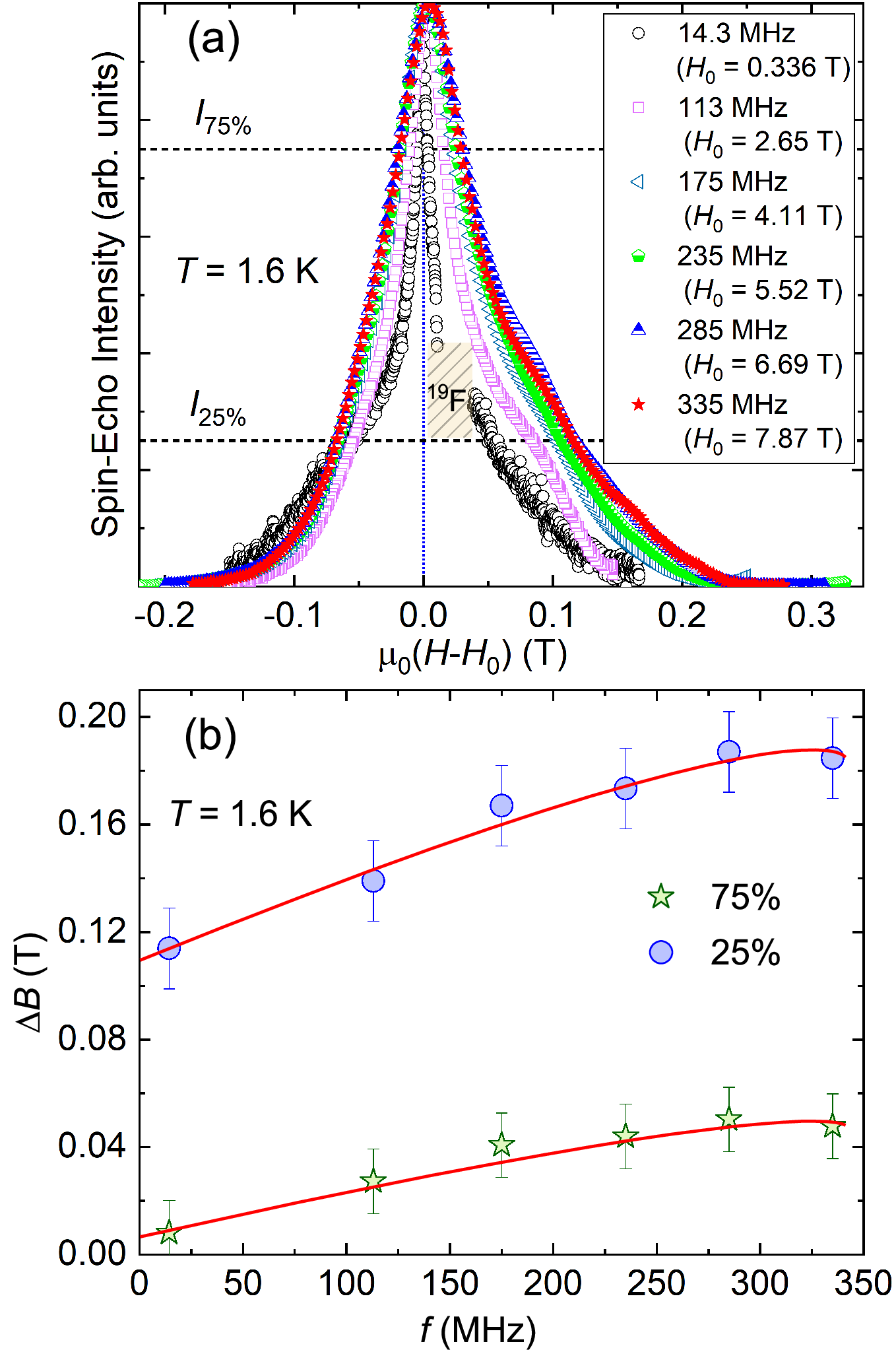}
\caption{(a) $^1$H NMR spectra recorded at $T = 1.6$~K for different frequencies (normalized to unity for comparison). The horizontal dashed lines correspond to 25$\%$ ($I_{25\%}$) and 75$\%$ ($I_{75\%}$) of the maximum intensity. The gap around  $\mu_0(H-H_0) = 0.025$~T  for $ f = 14.3$~MHz is due to the removal of strong signal from $^{19}$F NMR in our NMR probe.     
(b) The linewidth corresponding to $I_{25\%}$ ($\Delta H_{25\%}$) and $I_{75\%}$ ($\Delta H_{25\%}$) as a function of frequency.}
\label{Fig8}
\end{figure}
To further investigate the nature of magnetic ordering, we measured the $^1$H NMR spectra in different magnetic fields at $T = 1.6$~K well below $T^*$, across a range of NMR frequencies [see Fig.~\ref{Fig8}(a)]. At a low frequency of 14.3 MHz ($\sim$0.33 T), the spectrum can be divided into two components: one broad line and another narrow line. With increasing NMR frequency (or, magnetic field), the spectra broaden not only for the narrower but also for the broader component. To check the field dependence of the linewidth for both the components, we plotted the linewidth at 75\% ($\Delta B_{75\%}$) and 25\% ($\Delta B_{25\%}$) intensity positions as a function of the resonance frequency $f$  in Fig.~\ref{Fig8}(b). Both components increase with increasing resonance frequency and seem to saturate above $\sim$300 MHz. From the smooth extrapolation to $f = 0$ (or, $H = 0$), we got a finite linewidth of $\Delta B_{25\%} \simeq 0.115$~T, evidencing the spontaneous internal field originating from the LRO state. For the case of $\Delta B_{75\%}$, although $\Delta B_{75\%} \simeq 0.0114$~T at $H = 0$ is one order of magnitude smaller than $\Delta B_{25\%}$, the finite value of $\Delta B_{75\%}$ at $H = 0$ is also consistent with the LRO state. However, these results strongly suggest that not all protons experience the same internal field; rather, they are subject to different local magnetic environments depending on their positions within the lattice. Since the hyperfine coupling constants for H1 are much smaller than those of H2 and H3, the narrow line could be due to H1 site while the broad line may be originating from H2 and H3 sites. Moreover, it is often possible to infer the nature of the magnetic ordering from the shape of the NMR spectra in the ordered state~\cite{Ranjith014415}. However, in the present compound,
%due to weak hyperfine coupling of H with Cu$^{2+}$ ions, we were not able to trace the static internal field and hence, the $^1$H NMR line shape precisely. 
due to the superposition of three $^{1}$H lines, we were not able to determine the static internal field precisely and therefore, did not discuss the nature of the ordered state from the spectral shape.

\subsubsection{$^1$H spin-lattice relaxation rate $1/T_1$}
\begin{figure}
\includegraphics[width=\linewidth]{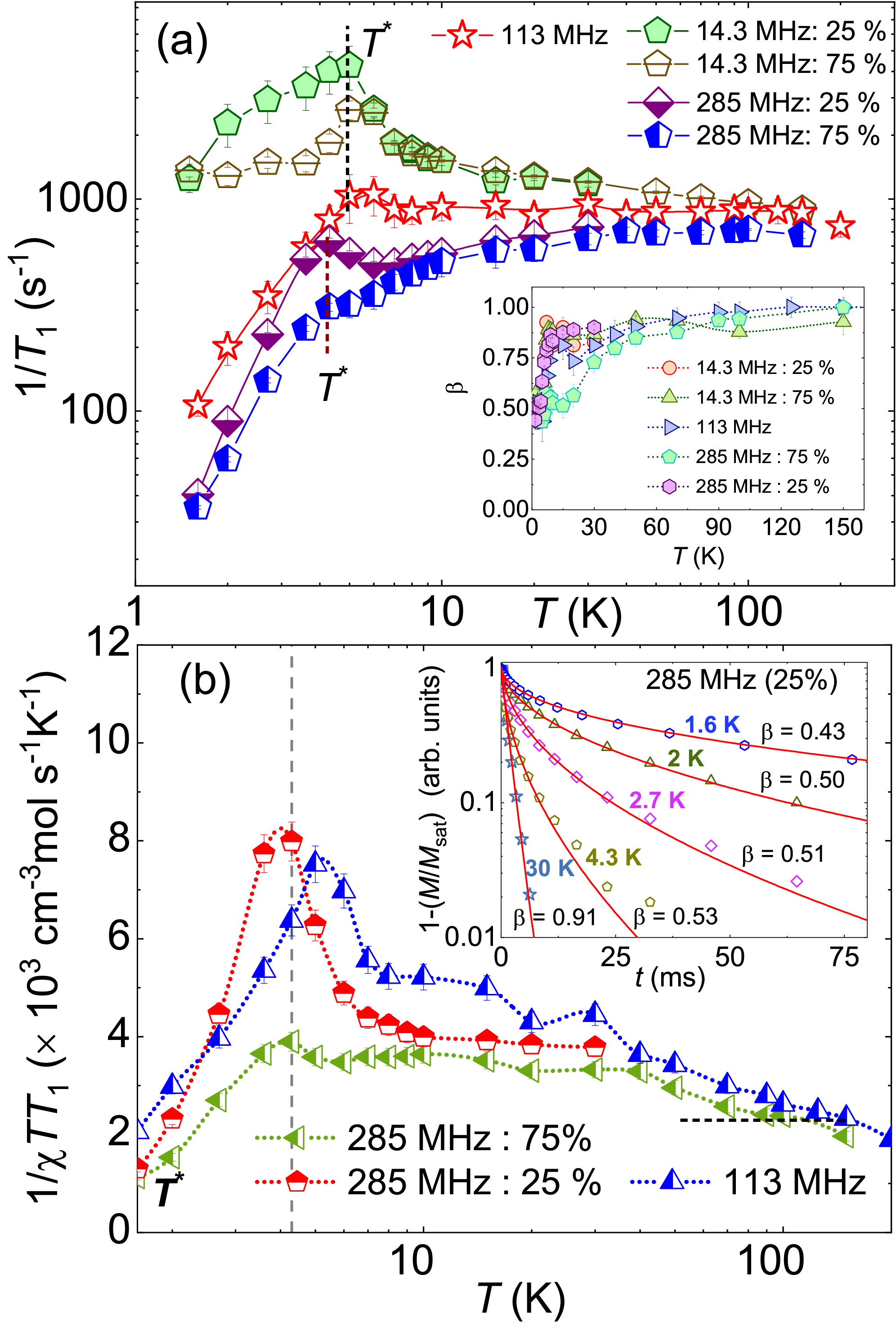}
\caption{(a) $^1$H spin-lattice relaxation rate ($1/T_1$) vs $T$ measured at different frequencies at the $I_{25\%}$ and $I_{75\%}$ positions from the high field side of the central peak. Solid lines are guides to the eyes. Inset: Temperature dependence of stretch exponent $\beta$. (b) $1/\chi T_1T$ vs $T$ at 285 and 113 MHz. The horizontal dashed line is a reference line for the temperature-independent behavior. Inset: Longitudinal magnetization recovery curves at different temperatures, and the solid lines are the fits using Eq.~\eqref{Eq11}.}
\label{Fig9}
\end{figure}
To investigate the local spin dynamics, we measured the $^1$H spin-lattice relaxation rate ($1/T_1$) at three different frequencies: 14.3~MHz (0.33~T), 113~MHz (2.65~T), and 285~MHz (6.69~T), down to the lowest temperature of 1.6~K. For 14.3~MHz and 285~MHz, $1/T_1$ was measured at two distinct positions on the spectrum corresponding to 75\% and 25\% of the maximum signal intensity at the high field side of the peak position. For 113~MHz, the measurement was performed only near the 75\% position of the central peak. 

The nuclear recovery curves for all measured frequencies were fitted with a stretched exponential function
\begin{equation}
1 - \frac{M(t)}{M(\infty)} = Ae ^{-(t/T_1)^\beta}.
\label{Eq11}
\end{equation}
Here, $M(t)$ is the nuclear magnetization at a time $t$ after the saturation pulse, $M(\infty$) is the equilibrium nuclear magnetization, and $\beta$ is the stretched exponent. The typical fitted recovery curves are shown in the inset of Fig.~\ref{Fig9}(b), and the corresponding $^1$H spin-lattice relaxation rates $1/T_1$, extracted from the fits, are presented in Fig.~\ref{Fig9}(a). As illustrated in the inset of Fig.~\ref{Fig9}(a), $\beta$ remains close to unity at high temperatures, indicating a uniform distribution of relaxation times. However, at low temperatures $\beta$ decreases significantly, pointing to a distribution of the relaxation rates, possibly due to the presence of some defects~\cite{Mohanty104424,Johnston176408,Ambika015803}.

At high temperatures, no obvious difference in $1/T_1$ is observed for the 75\% and 25\% of the intensity positions. However, a slight difference in magnitude observed at low temperatures can be attributed to the difference in hyperfine couplings for different H sites. The observed nearly temperature independent behaviour of $1/T_1$ at high temperatures for all the measured frequencies is consistent with the paramagnetic fluctuations. As the temperature is lowered, it consistently features a peak at $T^{*}$ for all frequencies and both the measured positions. This is a clear indication of the critical slowing down of spin fluctuations due to the onset of a magnetic LRO.
%For 14.3~MHz, a noticeable increase in $1/T_1$ is observed as the temperature is lowered, particularly below 10~K. This is due to the slowing down of spin fluctuations towards the ordered state, culminating in a weak peak at $T^{*} \simeq 5$~K (at $\mu_0 H = 0.33$~T), which signals the onset of a magnetic LRO.
This is in sharp contrast with the short-range type order observed in $C_{\rm p}(T)$. Thus, one possible explanation could be that the spin system orders at $T^*$ and a small fraction of defects still exists in the ordered state. These defects lead to a broadening of heat capacity peak and results in a Schottky anomaly. As compared to the peak ($T^{*} \simeq 5$~K) observed at 14.3~MHz, the peak position at 285~MHz is shifted slightly to low temperatures ($T^* \simeq 4.3$~K). This reflects that $T^*$ is suppressed with field, as typically expected for an AFM LRO~\cite{Cui174428,*Kikuchi224416}. This behavior of $T^*$ also ascertain that the shift of the broad maxima in $C_{\rm p}(T)$ to higher temperatures with magnetic field originates from the Schottky behaviour of the defects.

Below $T^*$, $1/T_1$ for both positions ($I_{25\%}$ and $I_{75\%}$) and all the measured frequencies decreases systematically, reflecting the magnon scattering (two-magnon or three magnon Raman) process in the ordered state. It is also noted that $1/T_1$ below $T^*$ shows a strong field dependency. In a low frequency (0.33~T), $1/T_1$ is weakly temperature dependent while in a higher frequency (6.7~T), it is reduced significantly with temperature. Such a rapid decrease in $1/T_1$ in higher fields could be due to the opening of a gap in the magnon spectrum~\cite{Beeman359,Belesi094422}. Since the magnon gap depends on the magnetic field directions, one may also expects a distribution of magnitude of the gap under magnetic field. Another possible scenario would be the presence of paramagnetic spin fluctuations associated with the defects in the ordered state, which get suppressed with the application of a magnetic field.
%{\color{blue} Such a behaviour can possibly be ascribed to strong quantum fluctuations persisting below $T^*$ which is suppressed with field.}
%\textbf{Such a behavior can be understood as the combined effect of (i) the opening of a Zeeman gap in the magnon spectrum, which suppresses Raman-type relaxation processes, and (ii) the reduction of low-energy spin fluctuations with increasing field. While the limited low-$T$ window in our data prevents a quantitative extraction of gap magnitude, the observed trend of $1/T_1(T,H)$ is qualitatively consistent with the Zeeman-gap scenario.}
Further studies including more systematic measurements of $1/T_1$ at much lower temperatures below 1.6~K and at various magnetic fields are required to have a better understanding of this behaviour.
%\textbf{It is also noted that $1/T_1$ below $T^*$ in a low magnetic field (0.33~T) shows a weak or nearly temperature independent behavior, suggesting temperature-independent magnetic fluctuations in the ordered state. Given the strong temperature dependence in $1/T_1$ observed under high magnetic fields, such nearly temperature-independent magnetic fluctuations are considered to be suppressed by the application of the magnetic field. Further studies including more systematic measurements of $1/T_1$ at much lower temperatures below 1.6~K and at various magnetic fields are required to understand the origin of this behaviour.

To visualize spin fluctuation effects in the paramagnetic regime above $T^*$, we examine the temperature dependence of $1/T_1T\chi$, as shown in Fig.~\ref{Fig9}(b). The quantity $1/T_1T$ is related to the imaginary part of the dynamic susceptibility, $\chi^{\prime\prime}_M(\vec{q},\omega_{\rm N})$, at the NMR frequency $\omega_{\rm N}$, via the following relation~\cite{Moriya23,*Moriya516}
\begin{equation}
\dfrac{1}{T_1T} = \dfrac{2\gamma^{2}_{\rm N}k_{\rm B}}{N^{2}_{\rm A}} \sum_{q}^{} |A(\vec{q})|^2 \dfrac{\chi^{\prime\prime} (\vec{q},\omega_{\rm N})}{\omega_{\rm N}},
\label{Eq12}
\end{equation}
where the summation is over wave vectors $\vec{q}$ in the first Brillouin zone, and $A(\vec{q})$ denotes the form factor of the hyperfine interaction. 
%\textbf{In the static limit, when the uniform fluctuations ($\vec{q} = 0$, $\omega = 0$) dominate, real part of the dynamic susceptibility $\chi'(0,0)$ corresponds to the uniform static susceptibility $\chi$ through the Kramers–Kronig relation~\cite{Slichter2013}
%\begin{equation}
%	\chi'(0) = \frac{2}{\pi} \int_0^\infty \frac{\chi^{\prime}(\omega^{\prime})}{\omega^{\prime}}\, d\omega^{\prime}.
%\end{equation}
%Thus, in the high-temperature paramagnetic regime, the ratio $1/T_1T\chi$ is expected to remain constant, as indicated by the dashed line in Fig.~\ref{Fig9}(b).}
%{\color{blue} However, upon lowering the temperature, we observed a significant enhancement in $1/T_1T\chi$ below approximately 75~K ($\sim 3\theta_{\rm CW}$). This enhancement signals the growth of AFM correlations with $\vec{q} \neq 0$, persisting well above $\theta_{\rm CW}$. Such extended spin correlations over a broad temperature range are characteristic of frustrated magnetic systems~\cite{Nath214430}.}
In the high-$T$ paramagnetic regime, where random spin fluctuations with no specific $\vec{q}$-dependence dominate, $1/T_1$ can be related to the real part of the dynamic susceptibility $\chi'(0,0)$ through the fluctuation–dissipation theorem~\cite{Sebastian104428}. Under the assumption that the autocorrelation function of the hyperfine field decays exponentially in time, one obtains the relation $1/T_1 \propto \chi T$~\cite{Slichter2013}. This leads to a temperature-independent $1/T_1T\chi$ behaviour, as indicated by the dashed line in Fig.~\ref{Fig9}(b). However, upon lowering the temperature, we observed a slight enhancement in $1/T_1T\chi$ below about 50~K which is of the order of $\theta_{\rm CW}$. This trend is likely due to the growth of AFM correlations with $\vec{q} \neq 0$ at low temperatures~\cite{Nath214430}.

\subsubsection{$^1$H spin-spin relaxation rate $1/T_2$}
\begin{figure}[h]
\includegraphics[width=\linewidth]{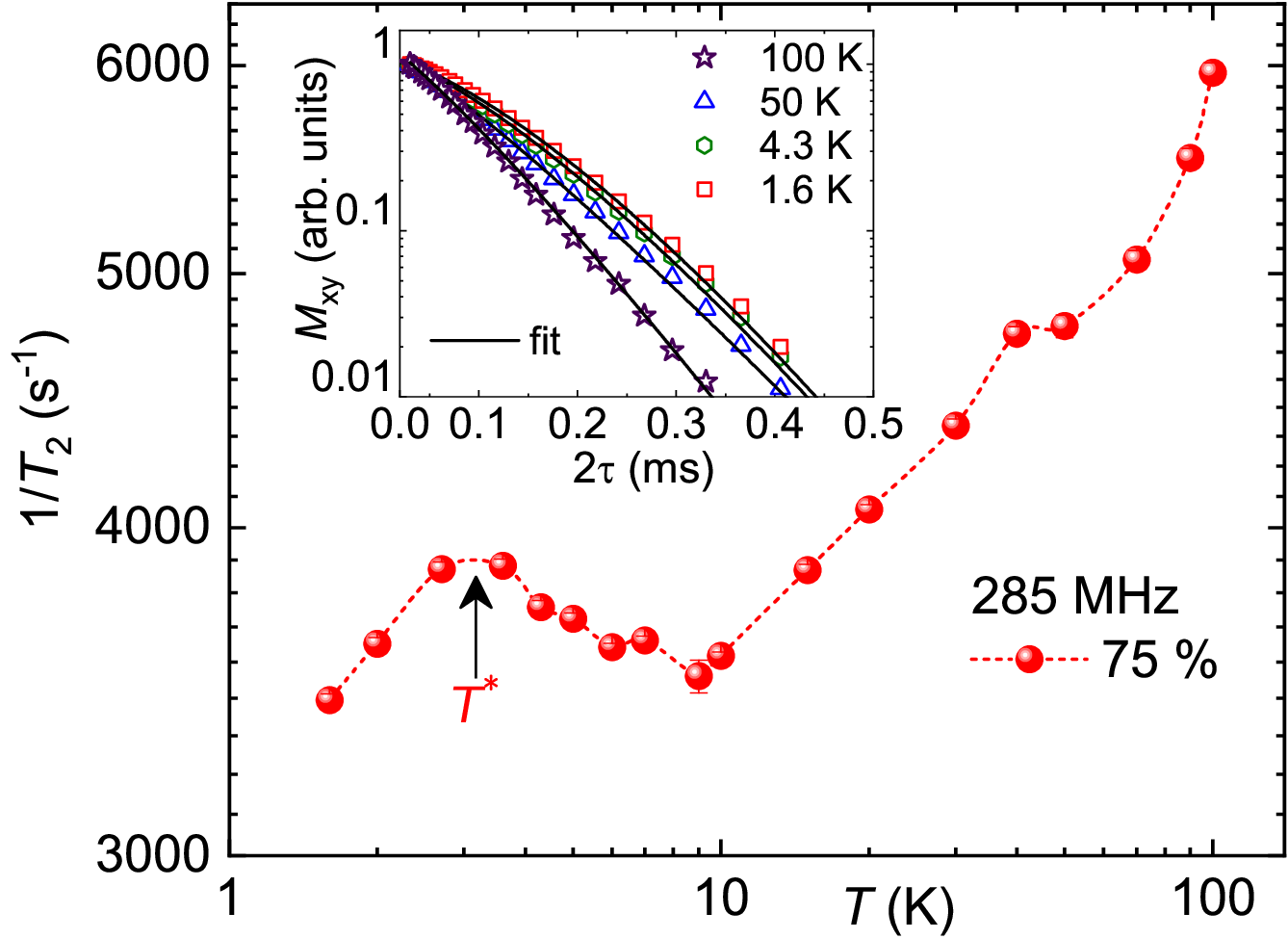}
\caption{$^1$H spin-spin relaxation rate ($1/T_2$) vs $T$ measured at 285~MHz and at the 75\% of the maximum intensity. Inset: Transverse magnetization decay curves at three selected temperatures and the solid lines are the fits using Eq.~\eqref{Eq13}.}
\label{Fig10}
\end{figure}
The $^1$H spin-spin relaxation rate ($1/T_2$) under a magnetic field of $\sim 6.69$~T was evaluated from the decay of the transverse magnetization ($M_{xy}$) as a function of the delay time $\tau$ between the $\pi/2$ and $\pi$ pulses~\cite{Slichter2013}. The decay curves were fitted using the stretched exponential function:
\begin{equation}
M_{xy} = M_0\, e^{-(2\tau/T_2)^\beta},
\label{Eq13}
\end{equation}
where $M_0$ is the initial transverse magnetization (see the inset of Fig.~\ref{Fig9}). In the paramagnetic regime, the decay follows a nearly Lorentzian form with $\beta \approx 1$, indicative of a homogeneous relaxation process. As the temperature is lowered, $\beta$ gradually increases, reaching approximately 1.4 at $T = 1.6$~K, suggesting a transition towards more Gaussian-like relaxation behavior. The temperature dependence of $1/T_2$ is shown in Fig.~\ref{Fig10}. With decreasing temperature, $1/T_2$ decreases steadily, and then exhibits a peak at around $T^*$, reflecting the onset of static spin correlations associated with the magnetic LRO.

\section{Summary}
The spin-1/2 magnet CTSOH features a frustrated depleted-kagome geometry and fosters captivate low temperature properties. Magnetic measurements suggest a possible magnetic LRO at $T^* \simeq 4$~K. This was further endorsed by the NMR relaxation ($1/T_1$ and $1/T_2$) measurements that manifest a clear anomaly at $T^*$. The nature of the ordering appears to be canted AFM type.

Surprisingly, the heat capacity data divulge a broad maximum at $T^*$ which moves towards high temperatures with field. This confirms the role of defects that broaden the heat capacity anomaly and induces Schottky-like behaviour.
%Although the heat capacity below $T^*$ is described by a power-law of the form $\gamma T^\alpha$, the reduced exponent $\alpha \simeq 1.2$ 
From the $\chi(T)$ and bond angle analysis, we infer the co-existence of AFM and FM interactions. The average nearest-neighbour AFM exchange coupling is estimated to be $J/k_{\rm B} \simeq 66$~K.
%Nevertheless, $1/T_1$ demonstrates persistent spin fluctuations down to 1.6~K and as well as up to high temperatures (well above $\theta_{\rm CW}$), reflecting highly frustrated nature of the compound.
The $^1$H NMR spectra reveal traits associated with three inequivalent H sites and from the $K$ vs $\chi$ plot, the corresponding hyperfine coupling constants are estimated. Though our findings on CTSOH establish a magnetic ordering at $T^*$, the exact nature of the ordering yet remains ambiguous and requires further experiments including neutron diffraction. Thus, CTSOH appears to be a promising system for exploring the emergent quantum effects arising due to geometric frustration and site depletion.

\acknowledgements
We would like to acknowledge SERB, India for financial support bearing sanction Grant No.~CRG/2022/000997 and DST-FIST with
Grant No.~SR/FST/PS-II/2018/54(C). SS also acknowledges Fulbright-Nehru Doctoral Research Fellowship Award No.~2997/FNDR/2024-2025 and the Prime Minister's Research Fellowship (PMRF) scheme, Government of India. Work at the Ames National Laboratory was supported by the U.S. Department of Energy, Office of Science, Basic Energy Sciences, Materials Sciences and Engineering Division. The Ames Laboratory is operated for the U.S. Department of Energy by Iowa State University under Contract No.~DEAC02-07CH11358.

%\bibliography{ref}

\end{document}